\documentclass[runningheads]{llncs}
\usepackage[T1]{fontenc}
\usepackage{graphicx}
\usepackage{lineno,hyperref}
\usepackage{amsmath}
\usepackage{txfonts}
\modulolinenumbers[5]

\hypersetup{pdfborder=0 0 0}

\begin{document}
%
\title{Can linear algebra create perfect knockoffs?}
%
%
\author{Christopher Hemmens \and Stephan Robert-Nicoud}
\authorrunning{C. Hemmens et al.}
%
\institute{School of Business and Engineering, University of Applied Sciences and Arts of Western Switzerland (HES-SO), 1400 Yverdon-les-Bains, Switzerland
\email{\{christopher.hemmens,stephan.robert\}@hes-so.ch}}
\maketitle              
\begin{abstract}
As new Model-X knockoff construction techniques are developed, primarily concerned with determining the correct conditional distribution from which to sample, we focus less on deriving the correct multivariate distribution and instead ask if ``perfect" knockoffs can be constructed using linear algebra. Using mean absolute correlation between knockoffs and features as a measure of quality, we do produce knockoffs that are pseudo-perfect, however, the optimization algorithm is computationally very expensive. We outline a series of methods to significantly reduce the computation time of the algorithm.
\
\keywords{Model-X Knockoffs, Linear Algebra}
\end{abstract}
\section{Introduction}

Candes et al. \cite{candes2018panning} asked how to control the false discovery rate in high-dimensional non-linear models, for example, what if you have more features than observations but you know only a handful of the features has any predictive power on the response variable? The authors' solution was the introduction of Model-X Knockoffs – a set of random variables that statistically mimic the original feature set but, by design, do not have any predictive power on the response variable. The definition of a Model-X Knockoff, $\tilde{X}$, for a feature set $X$ is as follows:

\begin{enumerate}
 \item for any subset $S\subset\{1,\dots,p\}$, $(X,\tilde{X})_{\text{swap}(S)}\stackrel{d}{=}(X,\tilde{X})$;
\item $\tilde{X}\Perp Y$ $|$ $X$ if there is a response $Y$.
\end{enumerate}

Note that as long as $\tilde{X}$ is constructed without knowledge of $Y$, condition 2 is automatically satisfied. \cite{candes2018panning} test this methodology by replicating \cite{wellcome2007genome}, who identify gene clusters that predict the development of Crohn's disease in patients. Whereas the original study was able to identify 9 gene clusters, by implementing the Model-X Knockoff methodology, \cite{candes2018panning} was able to identify 18 gene clusters. Since then, several steps have been taken to improve the methodology. For example, \cite{bates2021metropolized} add implementation of the Metropolis-Hastings (MH) algorithm (\cite{metropolis1953equation}, \cite{hastings1970monte}) which, although more computationally expensive, improves the ``quality" of the knockoffs as measured by the mean absolute correlation between the knockoffs and their features. The key insight with regard to adding the MH algorithm to the construction process is that it's a Markov chain method that is time-reversible, and therefore mimics the condition that features and knockoffs are able to be swapped without affecting the distribution. The one element that connects every approach to generating knockoffs since the 2018 paper is the focus on deriving the correct conditional distribution from which to sample the knockoffs. But is this a necessary step? In every case we start with a $p\times n$ matrix and our goal is to generate another $p\times n$ matrix that fulfils a number of linear constraints. Hence, it is possible to reframe this problem not as deriving the correct conditional distribution from which to sample, but a linear algebra optimization problem subject to a series of linear constraints. If we can ensure that the univariate distributions of the knockoffs match those of the features, which is trivial, and that the main co-moments between all the features and knockoffs match up, then for a relevant subset of multivariate distributions, we will have constructed the best possible knockoffs for that distribution. This process, however, is computationally very expensive, so one of the main aspects of this paper is a discussion on methods to reduce the computation time as much as possible while still ensuring pseudo-perfect knockoffs. In this paper, section 2 summarizes the literature on knockoff construction to date; section 3 outlines how to frame the problem using linear algebra; section 4 tests the methodology on a set of artificially generated random variables with enforced non-zero values of coskewness and cokurtosis; section 5 tests the methodology on a real dataset that in particular presents a serious challenge to the knockoff generation procedure in general; section 6 concludes the paper.

\section{Literature Review}

In addition to the initial knockoff construction algorithms put forth by \cite{candes2018panning} and \cite{bates2021metropolized}, a wide range of other knockoff-generating methods have been developed. For example, \cite{romano2020deep} use deep generative models to sample knockoffs using a criterion measuring the validity of the knockoffs produced by the model. \cite{berti2023new} use copulas to generate knockoffs. Essentially, if the distribution of the $p$ features is $F(x_1,\dots,x_p)$, then, by Sklar's Theorem (\cite{sklar1959fonctions}) there exists a $p$-copula $C$ and univariate distributions $F_1,\dots,F_p$ such that

$$F(x_1,\dots,x_p)=C\left[F_1(x_1),\dots,F_p(x_p)\right],\, \forall x_1,\dots,x_p$$

By carefully selecting a set of 2-copulas $D_1,\dots,D_p$, a new distribution can be created

$$H(x_1,\dots,x_p,x_{p+1},\dots,x_{2p})=C\left[D_1\left(F_1(x_1),F_1(x_{p+1})\right),\dots,D_p\left(F_p(x_p),F_p(x_{2p})\right)\right]$$

from which knockoffs can be constructed. \cite{kurz2022vine} also look at copulas but vine copulas. They use this method because it's especially strong for high-dimensionally modelling. However, unlike the method in \cite{berti2023new} that samples all the knockoffs at once, they generate the knockoffs recursively, like in \cite{candes2018panning} and \cite{bates2021metropolized}. Since the knockoff construction algorithm is typically random, \cite{ren2023derandomizing} introduce a method for derandomizing the procedure and making it more stable by constructing multiple sets of knockoffs and aggregating the results.

\newpage
\section{Methodology}

Suppose we have a multivariate distribution $$F(x_1,\dots,x_p)=Pr(X_1\le x_1,\dots,X_p\le x_p)$$ where the cumulative distribution function (CDF) of $X_i$ is $F_i:\mathbb{R}\rightarrow [0,1]$. Suppose also that we have a sample of this distribution with $n$ observations. Our goal is to use a linear algebra optimization approach to construct perfect Model-X knockoffs. That is, if the original sample is $\mathcal{X}\equiv(X_1,\dots,X_p)\in \mathbb{R}^{p\times n}$, then we want to find a matrix $\tilde{\mathcal{X}}\equiv(\tilde{X}_1,\dots,\tilde{X}_p)\in \mathbb{R}^{p\times n}$ such that the statistical moments of each feature and the other features are equal to the statistical moments of that feature's knockoff and the other features, as well as between that feature's knockoff and the other features' knockoffs, while minimizing the absolute value of the correlation between each feature and its knockoff. We focus on equating up to the fourth statistical moment: correlation, coskewness, and cokurtosis. To understand our approach, you have to know that generating knockoffs using linear algebra is disproportionately computationally expensive compared to sampling from a distribution and, as a result, our methodology concentrates on minimizing the amount of work a computer needs to do.

\subsection{Setting Constraints}

We define our (standardized) $(m,k)$-moment between variables $X$ and $Y$ to be 

$$\mu_{m,k}(X,Y) = \frac{\frac{1}{n}\sum_{i=1}^{n}\left((x_i-\mu_{1}(X))^k (y_i-\mu_{1}(Y))^{m-k}\right)}{(\mu_{2}(X))^{k/2}(\mu_{2}(Y))^{(m-k)/2}}$$


where $$\mu_{1}(X)=\frac{1}{n}\sum_{i=1}^n x_i $$  and  $$ \mu_{2}(X)=\frac{1}{n}\sum_{i=1}^n (x_i-\mu_{1}(X))^2$$


This means our first constraints on each knockoff are that they have the same distributions as their features. We ensure this by determining the distribution of the feature in advance, and then performing a Kolmogorov-Smirnov test on the knockoff. We set the constraint that the p-value of the test's result is as close to 1 as possible. The next set of constraints are simply that for every value of $i,j=1,\dots,p$ where $i\ne j$, and $2\le m\le 4$ with $1\le k\le (m-1)$, we have



\begin{enumerate}
	\item $\mu_{m,k}(\tilde{X}_i,X_j)=\mu_{m,k}(X_i,X_j)$
	\item $\mu_{m,k}(X_i,\tilde{X}_j)=\mu_{m,k}(X_i,X_j)$
	\item $\mu_{m,k}(\tilde{X}_i,\tilde{X}_j)=\mu_{m,k}(X_i,X_j)$
\end{enumerate}

We are careful not to duplicate any of these constraints, for example, $\mu_{4,1}\left(X_i,\tilde{X}_j\right)$ and $\mu_{4,3}\left(\tilde{X}_j,X_i\right)$ are the same condition, so we only include it once in the optimization process.

\subsection{Optimizing}

Now the constraints are set, we want to optimize the knockoffs such that the correlations between the features and their knockoffs are as close to zero as possible. In many circumstances, it may not be possible to generate knockoffs with zero correlation with their features, because the resultant multivariate distribution $(X,\tilde{X})$ still needs to have a covariance matrix that is positive-semidefinite in order to be a valid distribution. For this reason, we set the minimization value to be the sum of the squared correlations between the features and their knockoffs. Formally,

\begin{equation}
\sum_{i=1}^p \left(\mu_{2,1}\left(X_i,\tilde{X}_i\right)\right)^2
\label{eqn:corr}
\end{equation}

By performing the minimization in this way, we trend all the values to zero while also ensuring that the minimization of absolute correlations far from zero are prioritised over those already close to zero.

\subsection{Initial Guess}

Since this approach is computationally expensive, we want to start as close to a viable solution as possible before running the optimization. That way, if a single iteration of the optimization takes an excessively long time, we can reduce the parameter indicating the maximum number of iterations and be comfortable that we started with something approximating a viable knockoff. In this paper, we approximate $F(\cdot)$ using the Gaussian copula:

$$F(x_1,\dots,x_p)\approx C(x_1,\dots,x_p)\equiv\Phi\left(\Phi^{-1}(F_1(x_1)),\dots,\Phi^{-1}(F_p(x_p))\right)$$

This approximation allows use to generate an initial guess for the knockoffs that already minimizes (\ref{eqn:corr}) somewhat while matching the covariance constraints using the formula provided in Section 3.1.1 of \cite{candes2018panning}. All that remains is to set the model's tolerance, maximum iterations, and learning rate.

\section{Speed Test Results}

To test the efficiency of the algorithm, we randomly generate $n$ observations for $p$ independent normal random variables with mean 0 and variance 1. We then shuffle the elements to artificially create non-zero covariance, coskewness, and cokurtosis among the variables. For variable $X_i$,

\begin{itemize}
	\item[$\bullet$] if mod$(i,4)=1$, we weight smaller observations towards the end of the sample,
	\item[$\bullet$]  if mod$(i,4)=2$, we weight larger observations towards the end of the sample,
	\item[$\bullet$]  if mod$(i,4)=3$, we weight observations close to the mean towards the end of the sample,
	\item[$\bullet$]  if mod$(i,4)=0$, the sample is left unchanged.
\end{itemize}

We run the algorithm using a CPU 11th Gen Intel(R) Core(TM) i7-11700F @ 2.50GHz with 16GB RAM running on Windows 11 Home 64-bit OS.
%
%

\subsection*{\underline{$p=4$}}

Tables \ref{tbl:summ2} and \ref{tbl:corr2} show how the summary statistics of the knockoffs evolve as more iterations in the knockoff construction process are performed, in addition to the correlation between each feature and its knockoff, which we want to be as close to zero possible. Because of the way we construct our initial guess, these values are already very close to what we want the final result to be, so they don't vary much as the optimization process continues. The important question is whether the constraints on the moments are fulfilled and, if so, how quickly and how precisely.

\begin{table}
	\centering
\caption{Summary statistics for the knockoffs when $p=4$}	
\begin{tabular}{r | c | c c | c c | c c}
Knockoff & Iterations & \multicolumn{2}{c |}{Mean} & \multicolumn{2}{c |}{Variance} & \multicolumn{2}{c}{Dist. p-value} \\
 & $n=$ & 100 & 1000 & 100 & 1000 & 100 & 1000 \\
\hline
\hline
$\tilde{X}_1$ & Target & -0.1258 & -0.0020 & 0.8988 & 0.9247 & 100\% & 100\% \\
\hline
 & 3 & -0.1200 & -0.0070 & 0.9438 & 0.9440 & 60.4\% & 42.0\% \\
 & 10 & -0.1252 & -0.0021 & 0.9044 & 0.9244 & 98.6\% & 56.0\% \\
 & 20 & -0.1261 & -0.0021 & 0.8995 & 0.9247 & 99.2\% & 61.1\% \\
 \hline
 \hline
$\tilde{X}_2$ & Target & 0.0214 & -0.0529 & 0.8296 & 0.9997 & & \\
\hline
 & 3 & 0.0193 & -0.0534 & 0.9726 & 1.0224 & 75.8\% & 10.7\% \\
 & 10 & 0.0212 & -0.0529 & 0.8437 & 1.0019 & 65.0\% & 6.0\% \\
 & 20 & 0.0214 & -0.0529 & 0.8261 & 0.9995 & 76.5\% & 12.9\% \\
 \hline
 \hline
$\tilde{X}_3$ & Target & 0.0349 & -0.0494 & 0.8706 & 1.0176 & & \\
\hline
 & 3 & 0.0108 & -0.0561 & 0.9544 & 1.0388 & 89.1\% & 70.3\% \\
 & 10 & 0.0342 & -0.0494 & 0.8729 & 1.0192 & 50.1\% & 21.5\% \\
 & 20 & 0.0348 & -0.0494 & 0.8715 & 1.0175 & 81.9\% & 30.5\% \\
 \hline
 \hline
$\tilde{X}_4$ & Target & -0.0392 & 0.0327 & 0.9514 & 0.9751 & & \\
\hline
 & 3 & -0.0609 & 0.0309 & 1.0852 & 0.9778 & 59.2\% & 91.5\% \\
 & 10 & -0.0398 & 0.0327 & 0.9717 & 0.9757 & 42.4\% & 93.7\% \\
 & 20 & -0.0399 & 0.0327 & 0.9594 & 0.9751 & 35.8\% & 95.2\% \\
\end{tabular}
\label{tbl:summ2}
\end{table}

\begin{table}
	\centering
\caption{Correlation of each feature and its knockoff when $p=4$}	
\begin{tabular}{r | c | c c}
\text{Knockoff} & \text{Iterations} & \multicolumn{2}{c}{\text{Correlation with Feature}} \\
 & $n=$ & 100 & 1000 \\
\hline
$\tilde{X}_1$ & \text{Initial Guess} & 0.047 & 0.009 \\
 & 3 & 0.121 & 0.058 \\
 & 10 & 0.147 & 0.058 \\
 & 20 & 0.105 & 0.057 \\
\hline
$\tilde{X}_2$ & \text{Initial Guess} & -0.105 & -0.013 \\
 & 3 & -0.032 & 0.037 \\
 & 10 & 0.021 & 0.078 \\
 & 20 & 0.013 & 0.080 \\
\hline
$\tilde{X}_3$ & \text{Initial Guess} & 0.022 & -0.030 \\
 & 3 & 0.031 & -0.026 \\
 & 10 & 0.016 & -0.023 \\
 & 20 & 0.011 & -0.021 \\
\hline
$\tilde{X}_4$ & \text{Initial Guess} & -0.026 & -0.009 \\
 & 3 & 0.036 & -0.013 \\
 & 10 & 0.228 & -0.015 \\
 & 20 & 0.185 & -0.013 \\
\end{tabular}
\label{tbl:corr2}
\end{table}


\begin{figure}[h!]
  \centering
   \includegraphics[width=0.5\linewidth]{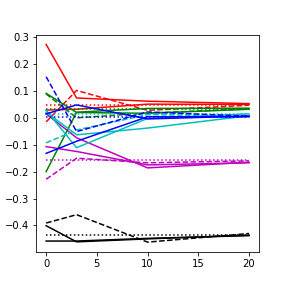}
   \caption{Evolution of \underline{correlation} constraint adherence over 20 iterations, $p=4$, $n=100$} 
   \label{fig:22corr}
\end{figure}

\begin{figure}[h!]
  \centering
   \includegraphics[width=0.7\linewidth]{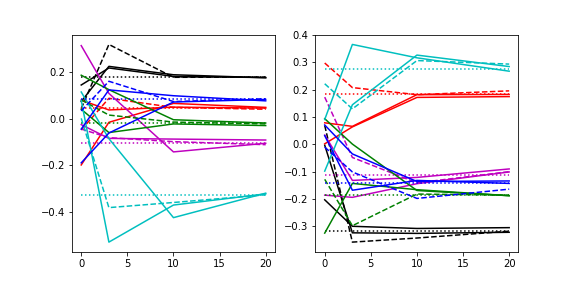}
   \caption{Evolution of \underline{coskewness} constraint adherence over 20 iterations, $p=4$, $n=100$} 
   \label{fig:22cosk}
\end{figure}

\begin{figure}[h!]
  \centering
   \includegraphics[width=\linewidth]{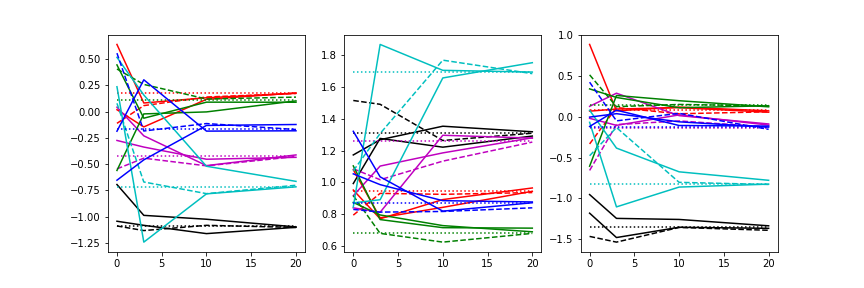}
   \caption{Evolution of \underline{cokurtosis} constraint adherence over 20 iterations, $p=4$, $n=100$} 
   \label{fig:22coku}
\end{figure}

\begin{figure}[h!]
  \centering
		\includegraphics[width=0.7\linewidth]{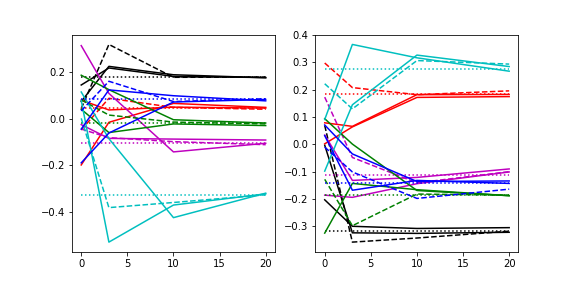}
		\caption{Coskewness Constraint, $p=4$, $n=100$ }
		\label{fig:coku_22}
\end{figure}

\begin{figure}[h!]
  \centering
		\includegraphics[width=0.7\linewidth]{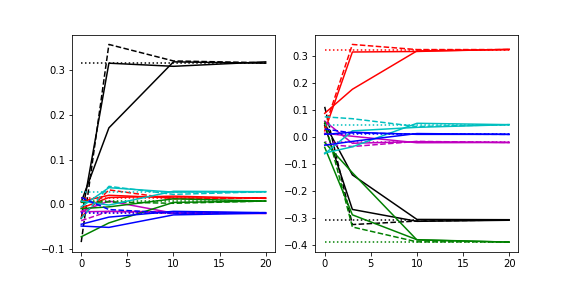}
		\caption{Coskewness Constraint, $p=4$, $n=1000$}
		\label{fig:coku_23}
\end{figure}

\begin{figure}[h!]
  \centering
		\includegraphics[width=0.7\linewidth]{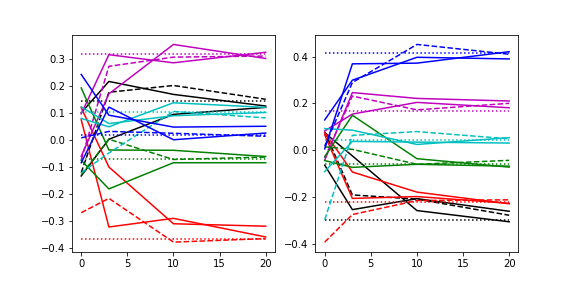}
		\caption{Coskewness Constraint, $p=8$, $n=100$ }
		\label{fig:coku_32}
\end{figure}

\begin{figure}[h!]
  \centering
		\includegraphics[width=0.7\linewidth]{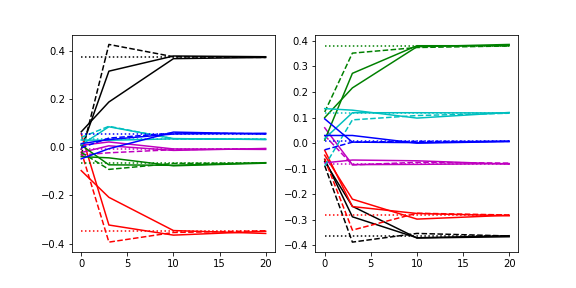}
		\caption{Coskewness Constraint, $p=8$, $n=1000$}
		\label{fig:coku_33}
\end{figure}

With four features, we have six relationships whose knockoff moments need to match those of their features. Figures \ref{fig:22corr}, \ref{fig:22cosk}, and \ref{fig:22coku} show the six relationships across the six moments of interest: correlation, left and right coskewness, and left, central, and right cokurtosis, and their evolution as we perform additional iterations in the optimization process.  Each colour represents a different relationship, for example, black shows the relationship between the first and second features, while green shows the relationship between the second and third features. The dotted line shows the real value of the relationship and therefore the target value of the constraint. The two solid lines show the value of the moment between one of the features and the knockoff of the other feature, while the dashed line shows the value of the moment between the two knockoffs. We can see that the optimization process is reaching the target moments quickly and precisely with this small dataset so we can be confident that the technique is working and producing close to pseudo-perfect knockoffs. Figure \ref{fig:coku_22}-\ref{fig:coku_33} show that if we increase the number of features, we see some degradation in the speed with which the constraints converge. In this context, it's important to note that going from 4 features to 8 increases the number of relationships from 6 to 28. In the current code, all of the cokurtosis constraints, for example, are entered into the model as a single constraint. The same is true for coskewness and correlation. This is to make the code as computationally efficient as possible, however, we could make the constraints for left coskewness and right coskewness separate constraints to make them converge sooner. The question is whether the increase in speed of convergence is adequately balanced with the loss of computational efficiency. In Figures \ref{fig:coku_22}, and  \ref{fig:coku_23} where $p=4$, we show all the relationships. In \ref{fig:coku_32} and \ref{fig:coku_33}, where $p=8$, we just show the quintiles for the sake of readability.

\vspace{3mm}

In terms of speed, Table \ref{tbl:time} shows how long each optimization process took. The code is already optimized to scale well with the number of features, $p$. We hope to find additional efficiency savings for the total number of observations.

\begin{table}
\begin{center}
\caption{Time taken to complete optimization for a set number of iterations in H:M:S}
	\begin{tabular}{r r r | c c c}
		\multicolumn{3}{c |}{Total Observations} & \multicolumn{3}{c}{Iterations} \\
		\hline
		$n$ & $p$ & $n\times p$ & 3 & 10 & 20 \\
		\hline
		100 & 4 & 400 & 0:00:07 & 0:00:13 & 0:00:26 \\
		 & 8 & 800 & 0:00:26 & 0:00:52 & 0:01:27 \\
		 \hline
		1000 & 4 & 4000 & 0:01:37 & 0:06:21 & 0:12:40 \\
		 & 8 & 8000 & 0:10:03 & 0:41:37 & 1:26:15 \\
	\end{tabular}
\label{tbl:time}
\end{center}
\end{table}

\begin{table}[h!]
\centering
\caption{Evolution of constraint values where 0 is the target}
\begin{tabular}{r | c | r}
Moment & Iterations & Constraint Value \\
\hline
Correlation & Initial Guess & 0.825 \\
& 20 & 0.049 \\
& 50 & 0.007 \\
& 100 & 0.006 \\
\hline
Coskewness & Initial Guess & 38.106 \\
& 20 & 1.031 \\
& 50 & 0.117 \\
& 100 & 0.090 \\
\hline
Cokurtosis & Initial Guess & 1692.465 \\
& 20 & 22.992 \\
& 50 & 2.477 \\
& 100 & 1.757 \\
\end{tabular}
\label{table:realcons}
\end{table}

\section{Real Data Results}
To test the effectiveness of this approach on real data, we use daily returns from the constituents of the Swiss Market Index (SMI) from the 26th May 2017 until the 8th April 2019. This is 481 observations from 20 features. The daily returns of these 20 stocks are scaled to have mean 0 and variance 1 and the scaled dataset is fit to a Gaussian copula. All univariate variables are modelled as Student-t distributions for which the degrees of freedom range from 3.08 to 5.81. Using the copula, we generate the initial guess and perform the optimisation procedure as outlined in Section 3. Since there are 20 features, we have 190 relationships and 1140 moment-based constraints. Table \ref{table:realcons} shows the aggregated constraints over a given number of iterations. The target for each constraint is zero. The time to complete 50 iterations was 6:45:56. The goal of the procedure is to produce knockoffs whose correlations with their features are as close to zero as possible whilst adhering to the moment constraints, and Table \ref{table:corrs} shows the quartiles of the correlations achieved. Table \ref{table:Ionn} shows the results for Credit Suisse and UBS, which got the lowest and highest resultant correlations respectively, as well as Roche, which is a big company that is representative of the middle of the distribution with a correlation of 0.54.

\begin{table}[h!]
\centering
\caption{Minimization results for correlation between knockoffs and features}
\begin{tabular}{r | c c c c c}
Quartile & Min & 0.25 & 0.5 & 0.75 & Max \\
\hline
Correlation & 0.27 & 0.47 & 0.56 & 0.69 & 0.88
\end{tabular}
\label{table:corrs}
\end{table} 

\begin{table}[h!]
\centering
\caption{Adherence of univariate distribution constraints}
\begin{tabular}{r | c c c c c}
Quartile & Min & 0.25 & 0.5 & 0.75 & Max \\
\hline
Feature KS p-value & 0.35 & 0.58 & 0.64 & 0.72 & 0.96 \\
Initial Guess KS p-value & 0.02 & 0.34 & 0.54 & 0.81 & 0.98 \\
Knockoff KS p-value & 0.02 & 0.45 & 0.80 & 0.87 & 0.98 \\
\end{tabular}
\label{table:ks}
\end{table} 

\begin{table}[h!]
\centering
  \caption{Resultant correlations between UBS, Roche and Credit Suisse}
 \begin{tabular}{r r | c | c c || c | c c}
 & &  \multicolumn{3}{c |}{Roche} & \multicolumn{3}{c}{UBS} \\
 & & Real & Feature & Knockoff & Real & Feature & Knockoff \\
 \hline
 Credit Suisse & Correlation & 0.190 & 0.191 & 0.190 & 0.625 & 0.618 & 0.618 \\
 \hline
 & Left Coskewness & -0.128 & -0.135 & -0.141 & -0.283 & -0.282 & -0.269 \\
 & Right Coskewness & -0.130 & -0.123 & -0.135 & -0.131 & -0.141 & -0.135 \\
 \hline
 & Left Cokurtosis & 1.055 & 0.996 & 1.050 & 2.635 & 2.574 & 2.661 \\
 & Central Cokurtosis & 1.491 & 1.509 & 1.493 & 2.369 & 2.298 & 2.338 \\
 & Right Cokurtosis & 1.549 & 1.533 & 1.549 & 2.349 & 2.423 & 2.367 \\
 \hline
 \hline
 Roche & Correlation & & & & 0.337 & 0.331 & 0.333 \\
 \hline
 & Left Coskewness & & & & -0.056 & -0.045 & -0.050 \\
 & Right Coskewness & & & & -0.108 & -0.099 & -0.111 \\
 \hline
 & Left Cokurtosis & & & & 1.504 & 1.588 & 1.522 \\
 & Central Cokurtosis & & & & 1.744 & 1.729 & 1.733 \\
 & Right Cokurtosis & & & & 1.558 & 1.570 & 1.585 \\
 \end{tabular}
 \label{table:Ionn}
 \end{table}
 
 
 \section{Discussion}
 If one uses mean absolute correlation in order to determine the ``quality" of a knockoff, then it seems that this algorithm is indeed capable of generating ``perfect" knockoffs. The question is how much time is required in order to generate them. By using the Gaussian copula to estimate the multivariate distribution of the features, we're able to generate an initial guess for the model in a matter of seconds that not only gives us knockoffs whose distributions already match those of the features, but also already satisfy to some extent the correlation requirements of good-quality knockoffs.  With this initial guess, the minimization algorithm doesn't need to run for that many iterations in order align the knockoffs with correct multivariate distribution as well as the constraints of the higher-order moments: namely coskewness and cokurtosis. As an initial step, this is a very promising technique for generating Model-X knockoffs, but the nature of the optimization algorithm suggests that there is lots of scope to significantly improve the computation cost of this approach, thereby significantly increasing the accessibility to ``perfect" knockoffs for a wider range of problems.

\bibliography{knockoff}

 \section{Notice}
 {\it This preprint has not undergone peer review or any post-submission
improvements or corrections. The Version of Record of this contribution is
published in "Big Data and Internet of Things. BDIoT 2024". Lecture
Notes in Networks and Systems, vol 887. Springer, Cham, and is available
online at https://doi.org/10.1007/978-3-031-74491-4\_81}
 
\end{document}